\documentclass[%
 reprint,
superscriptaddress,
 preprintnumbers,
 nofootinbib,
 nobibnotes,
 bibnotes,
 amsmath,amssymb,
 aps,
 floatfix,
]{revtex4-1}

\usepackage{graphicx}
\usepackage{dcolumn}
\usepackage{bm}
\usepackage{comment}
\usepackage{color}
\usepackage{natbib}
\usepackage[nowatermark]{fixmetodonotes}
\usepackage{isotope}
\usepackage{verbatim}

\usepackage{siunitx}
\DeclareSIUnit[number-unit-product = ]\percent{\char`\%}

\usepackage{hyperref}
\usepackage[mathlines]{lineno}

\begin{document}


\title{Measurement of the neutron capture cross section on argon} 

\author{V. Fischer}
\affiliation{University of California at Davis, Department of Physics, Davis, CA 95616, U.S.A.}

\author{L. Pagani}
\email{lpagani@ucdavis.edu}
\affiliation{University of California at Davis, Department of Physics, Davis, CA 95616, U.S.A.}

\author{L. Pickard}
\affiliation{University of California at Davis, Department of Physics, Davis, CA 95616, U.S.A.}

\author{A. Couture}
\affiliation{Los Alamos National Laboratory, LANSCE, Los Alamos, NM 87545, U.S.A.}

\author{S. Gardiner}
\affiliation{University of California at Davis, Department of Physics, Davis, CA 95616, U.S.A.}

\author{C. Grant}
\affiliation{Boston University, Department of Physics, Boston, MA 02215, U.S.A.}

\author{J. He}
\affiliation{University of California at Davis, Department of Physics, Davis, CA 95616, U.S.A.}

\author{T. Johnson}
\affiliation{University of California at Davis, Department of Physics, Davis, CA 95616, U.S.A.}

\author{E. Pantic}
\affiliation{University of California at Davis, Department of Physics, Davis, CA 95616, U.S.A.}

\author{C. Prokop}
\affiliation{Los Alamos National Laboratory, LANSCE, Los Alamos, NM 87545, U.S.A.}

\author{R. Svoboda}
\affiliation{University of California at Davis, Department of Physics, Davis, CA 95616, U.S.A.}

\author{J. Ullmann}
\affiliation{Los Alamos National Laboratory, LANSCE, Los Alamos, NM 87545, U.S.A.}

\author{J. Wang}
\affiliation{University of California at Davis, Department of Physics, Davis, CA 95616, U.S.A.}

\collaboration{ACED Collaboration}

\date{\today}

\begin{abstract}
The use of argon as a detection and shielding medium for neutrino and dark matter experiments has made the precise knowledge of the cross section for neutron capture on argon an important design and operational parameter. 
Since previous measurements were averaged over thermal spectra and have significant disagreements, a differential measurement has been performed using a Time-Of-Flight neutron beam and a $\sim$4$\pi$ gamma spectrometer. 
A fit to the differential cross section from $0.015-0.15$\,eV, assuming a $1/v$ energy dependence, yields $\sigma^{2200} = 673 \pm 26 \text{ (stat.)} \pm 59 \text{ (sys.)}$\,mb.
\end{abstract}

\pacs{Valid PACS appear here}

\maketitle



\section{\label{sec:introduction}Introduction}

Argon is a common detection medium used in many particle physics experiments. 
As a noble element, it has a low affinity for electron absorption and can therefore be used in Time Projection Chambers (TPC) or other applications where long distance electron or ion drift is desirable. 
Due to its low cost compared to other noble elements, such as xenon or neon, it has been used in very large neutrino detectors including ICARUS~\cite{Antonello:2013ypa}, MicroBooNE~\cite{Acciarri:2016smi}, and protoDUNE~\cite{Abi:2017aow}. 
A large liquid argon TPC is also planned for the DUNE experiment~\cite{Acciarri:2016ooe}.
Argon is also an excellent scintillator and can be made very radiologically clean, thus it is also used as a target in dark matter experiments, for instance DarkSide~\cite{Aalseth:2017fik}, and as a shield for neutrinoless double beta decay experiments such as GERDA~\cite{Ackermann:2012xja} and LEGEND~\cite{Abgrall:2017syy}.

All these applications rely on having a complete understanding of the transport of neutrons through liquid argon and the physics of the (n,$\gamma$) capture process on natural argon. 
There has been only three measurements of the thermal-neutron capture cross section, and these have yielded inconsistent results~\cite{Koehler:196318a, French:1965225, RANAKUMAR1969333}. 
All were done by activating samples of argon in a nuclear reactor, counting the beta decay of the $^{41}$Ar daughter in a gamma spectrometer, and then making appropriate corrections to convert the reactor spectral averaged cross section to the standard thermal cross section ($\sigma^{2200}$). 
In this paper we report the results of a differential measurement of the neutron capture cross section on argon in the thermal energy range using the Detector for Advanced Neutron Capture Experiment (DANCE) at Los Alamos National Laboratory.


\section{\label{sec:setup}EXPERIMENTAL SETUP}

DANCE is located on Flight Path\,14 at the Lujan Neutron Scattering Center~\cite{LANSCEBeam:1990} at a distance of $20.25$\,m from the upper-tier water moderator of the accelerator-driven pulsed neutron source.
The neutrons are produced via spallation reactions caused by an $800$\,MeV proton beam, impinging on a tungsten target, with a typical beam current of $80$\,$\mu$A.
A mercury shutter can be used to prevent neutrons reaching the target, allowing different run types to be performed as described in Sec.~\ref{sec:selection}.
The gamma detector is a nearly 4$\pi$ gamma ray calorimeter composed of a spherical array of 160\,BaF$_2$ crystals, each with a volume of $734$\,cm$^3$ and monitored by a photomultiplier tube (PMT). 
Adjacent to the crystals' inner face and surrounding the evacuated beam pipe where neutrons are traveling from the moderator, a $6$\,cm thick $^6$LiH shell has been installed to attenuate the rate of scattered neutrons capturing on the BaF$_2$ crystals. 
A detailed description of the DANCE setup can be found in Refs.~\cite{Heil:2013oua,MOSBY201475}.
This Time-Of-Flight (TOF) system is capable of measuring neutron energy in the thermal range with a few meV accuracy.

The Argon Capture Experiment at DANCE (ACED) consists of a target volume of argon gas located at the center of the DANCE spectrometer.
This target consists of a hollow aluminum cylinder that is $2.9$\,cm long, $3$\,cm in diameter, sealed by two $0.0762$\,mm thick Kapton windows spaced apart by $2.30 \pm 0.05$\,cm. 
This uncertainty is estimated from the dimensional tolerances specified on the target blueprints.
These Kapton windows allow neutrons to pass through while minimizing scattering. 
The same apparatus was used for a previous measurement of the $^{136}$Xe neutron capture cross section~\cite{PhysRevC.94.034617}. 
The target volume was filled with high purity ($99.999$\,\% Ar) gaseous natural argon and pressurized above the local atmospheric pressure throughout the data taking period. 
As described in Sec.~\ref{sec:strategy}, background measurements were taken with the target fully evacuated and kept at the same vacuum level as the surrounding beam pipe.

\begin{figure}[tpb]
\centering
\includegraphics[width=\columnwidth]{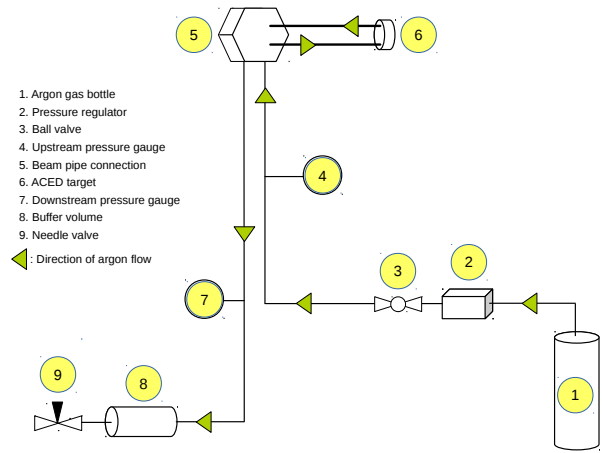}
\caption{Schematic of the ACED gas system.}
\label{fig:gas_system_schematic}
\end{figure}

A detailed schematic of the ACED gas system is displayed in Fig.~\ref{fig:gas_system_schematic}.
An electronic regulator was used to control the absolute pressure in the target, which was monitored, along with the temperature, by two temperature compensated gauges (Additel 681-02) located upstream and downstream of the vessel.
A buffer volume was used to smooth out small pressure fluctuations in the system.
As a result, during the data taking periods the average pressure and temperature of argon were $1.0987 \pm 0.0005$\,bar absolute and $296.7 \pm 1.0$\,K, respectively.
These values yielded an average density of $\rho = (1.779 \pm 0.006) \times 10^{-3}$\,g/cm${^3}$.
The uncertainty in temperature was derived from the observed temperature fluctuation in the experimental hall.
The error on the average density takes into account the intrinsic accuracy of the gauges, the difference between the upstream and downstream gauges, and the stability of the pressure and temperature over the duration of each run.



\section{\label{sec:strategy}MEASUREMENT STRATEGY and RUN DESCRIPTION}

Events in DANCE are recorded in a $15$\,ms acquisition window triggered by either a beam spill or an external pulser - provided the energy deposit is above $150$\,keV.

Data undergo several stages of processing. 
Initially the intrinsic alpha-induced background present in the BaF$_2$ crystals\footnote{The intrinsic radioactivity of BaF$_2$ originates from the $\alpha$-decay chain of the chemical homologue $^{226}$Ra (typically $0.2$\,Bq/cm$^{3}$~\cite{Heil:2013oua}). Most of the DANCE crystals have an intrinsic alpha activity ranging between $150$ and $250$\,Hz.} is removed on a crystal-by-crystal basis by applying a time profile discrimination cut. 
This selection is based on the difference in intensity ratio between the fast and slow components of the BaF$_2$ scintillation light~\cite{Reifarth:2013xny} as a function of the radiation type in this case, between alphas and gammas.
Next, individual gamma rays are reconstructed via a nearest neighbor clustering algorithm, in which all adjacent crystals, contributing to a single event, are grouped into a cluster. 
The number of such clusters gives the reconstructed cluster multiplicity of that event; the energy of the cluster provides the individual reconstructed gamma energy, with the summed energy of all clusters yielding the total reconstructed event energy. 
The need for such an algorithm is necessitated by gammas undergoing multiple Compton scatters and depositing energy in adjacent crystals.

The ACED data taking period ran from the 2nd to the 14th November 2018 and was split into 6 distinct non-calibration run types intermittently spaced throughout this period.
Each run type was designed to allow a full understanding of the detector conditions, response, and backgrounds.
This included four ``beam on'' run types where the target was either filled with argon, or completely evacuated, and with the beam shutter open or closed.
Additionally, there were two ``beam off'' runs in which the target was either filled with argon, or completed evacuated.


\section{\label{sec:selection}RUN SELECTION and CALIBRATION}

Data quality cuts (e.g. run duration, expected crystal occupancy, sufficient number of alpha events to perform calibration) are applied to ensure good detector performance. 
This reduces the overall data-set by about $25$\,\%. 
Furthermore, of the 160 crystals three did not show satisfactory results due to their low gain, hence were removed from the analysis.

The energy calibration of DANCE, using $^{22}$Na, $^{88}$Y, $^{60}$Co, and $^{239}$Pu-$^{9}$Be radioactive sources, was performed before and after the argon target runs.
Each crystal's energy response was found to be linear to within $2.4$\,\%. 
The energy scale is accurate to within $1.4$\,\% in the energy range of interest ($<6$\,MeV). 
This was deduced by comparing the reconstructed to nominal energies of the calibration sources.
Moreover, the stability of the energy scale throughout the data taking period was assessed. 
This was achieved by looking at the energy deposited by the most energetic, fully-contained alpha from the intrinsic radium contamination, and was found stable to within $1$\,\%.

A total energy cut is applied in order to select events from neutron captures on $^{40}$Ar.
Given the Q-value of $6.0989$\,MeV, events need to deposit a total energy of $6.1^{+0.5}_{-0.9}$\,MeV in the detector. 
Here, the energy window accounts for: the $150$\,keV hardware threshold, the energy scale uncertainty, and the $4$\,\% resolution of the detector at this energy (measured using the calibration data).
In addition, we require more than one reconstructed cluster - this is motivated by the fact that a neutron capture on an s-state for a $0^+$ nuclei will produce multiple gammas.
This is confirmed in the CapGam database~\cite{Hardell:1970tncna, Nesaraja:2016ktw}.
Moreover, requiring multiple clusters in an event suppresses the constant-in-time (CIT) background due to natural radioactivity within the crystals and from outside of the detector~\cite{Reifarth:2013xny}.

To assess the efficiency of measuring the $^{41}$Ar gamma cascade, a Geant4-based~\cite{Agostinelli:2003250, Allison:2016186, Jandel:2007ge} simulation was performed. 
This modelled, on a crystal-by-crystal basis, the energy resolution and the minimum detectable deposited energy - along with their uncertainties ($0.6$\,\%). 
Furthermore, it accounted for the uncertainty in the intensities of the major gamma lines~\cite{Hardell:1970tncna, Nesaraja:2016ktw} in the $^{41}$Ar gamma cascade ($0.7$\,\%).
Finally, it included the combined effect of the Q-value and cluster multiplicity cuts.
This approach yielded a result of $\varepsilon = 98.9 \pm 0.3 \text{ (stat.)} \pm 0.9 \text{ (sys.)}$\,\%.

Two beam monitors, $^{6}$Li and $^{3}$He, located downstream of the target volume, were used to understand the neutron beam flux as a function of energy.
These utilize the known neutron capture cross sections for the $^{6}$Li(n,$\alpha$)$^{3}$H and $^{3}$He(n,p)$^{3}$H interactions, respectively. 
This, combined with the neutron beam size being smaller than both the argon target volume and the beam monitors, allowed the determination of the total neutron flux.
In this analysis, we used the $^{6}$Li monitor to measure the beam flux with the $^{3}$He monitor providing an essential consistency check.
The absolute calibration of the beam monitor was obtained using the controlled activation of a sodium sample, which was placed in the neutron beam at the same location as the argon target volume.
The activated sodium target was later counted on a HPGe detector to determine the neutron flux at DANCE.
Using the energy profile of the beam monitors, only thermal neutrons ($<0.2$\,eV) were selected and their contribution to the total flux was calculated.
A detailed description of this method can be found in Ref.~\cite{SodiumPaper}.
This procedure has an uncertainty of $5.8$\,\% when combined with the detector stability over time and is the dominant uncertainty in the experiment (see Tab.~\ref{tab:xsec_sys_errors}).

\begin{figure}[tb]
\centering
\includegraphics[width=\columnwidth]{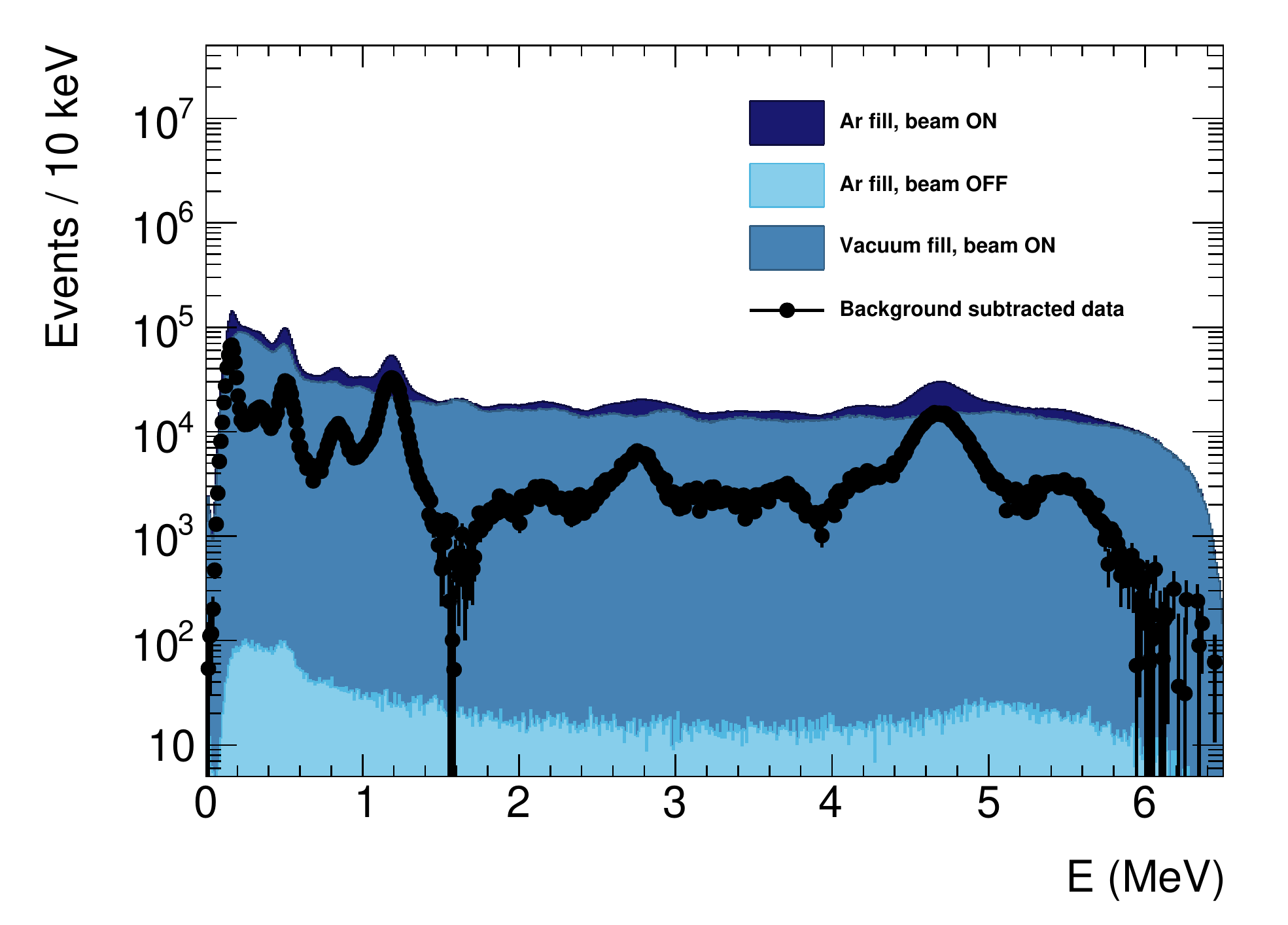}
\caption{Energy spectra of individual clusters for the different ACED data-sets. 
Only events detected in the $0.02-0.04$\,eV neutron energy window, and satisfying both Q-value and cluster multiplicity cuts are selected.}
\label{fig:bkg_subtraction}
\end{figure}

In order to quantify events coming from neutron captures on argon it is necessary to perform a background subtraction. 
To do this the following three data-sets are used: beam incident on the argon filled target (A), no beam incident on the argon filled target (S), and beam incident on the target under vacuum (V).

In the following text, $T_{0}$ refers to the number of beam spills, $D$ is the number of events, $R$ is the rate of events, $\Phi$ is the neutron flux, and $\sigma_{a,b}$ are the neutron capture cross sections on argon and on the sum of surrounding materials, respectively. 
The number of CIT background events are measured in data-set S, given by:
\begin{equation}
\label{eq:ds}
D^{S} = T_{0}^{S} R.
\end{equation}

Analogously, in data-set V, $D^{V}$ is the sum of the background events due to the flux of scattered neutrons ($\Phi^V$) captured in the materials surrounding the target (e.g. the Kapton windows) together with CIT backgrounds:
\begin{equation}
\label{eq:dv}
D^{V} = T_{0}^{V} R + \Phi^{V} \sigma_b.
\end{equation}

Furthermore, in data-set A, $D^{A}$ is comprised of CIT backgrounds, beam-related gammas passing through the shielding, background events due to captures on the surrounding materials, and neutron captures on $^{40}$Ar:
\begin{equation}
\label{eq:da}
D^{A} = T_{0}^{A} R + \Phi^{A} \sigma_b + \Phi^{A} \sigma_a.
\end{equation}

The number of neutron captures on argon can then be calculated using Eqs.~\ref{eq:ds} to \ref{eq:da}:
\begin{equation}
\label{eq:nn}
\Phi^{A} \sigma_a = D^{A} - \frac{T_{0}^{A}}{T_{0}^{S}} D^{S} - \frac{\Phi^{A}}{\Phi^{V}} \left(D^V-\frac{T_{0}^{V}}{T_{0}^{S}} D^{S}\right).
\end{equation}

Fig.~\ref{fig:bkg_subtraction} shows an example of this procedure, together with the energy spectra of the different ACED data-sets.


\section{\label{sec:result}RESULTS}

\begin{figure}[tb]
\centering
\includegraphics[width=\columnwidth]{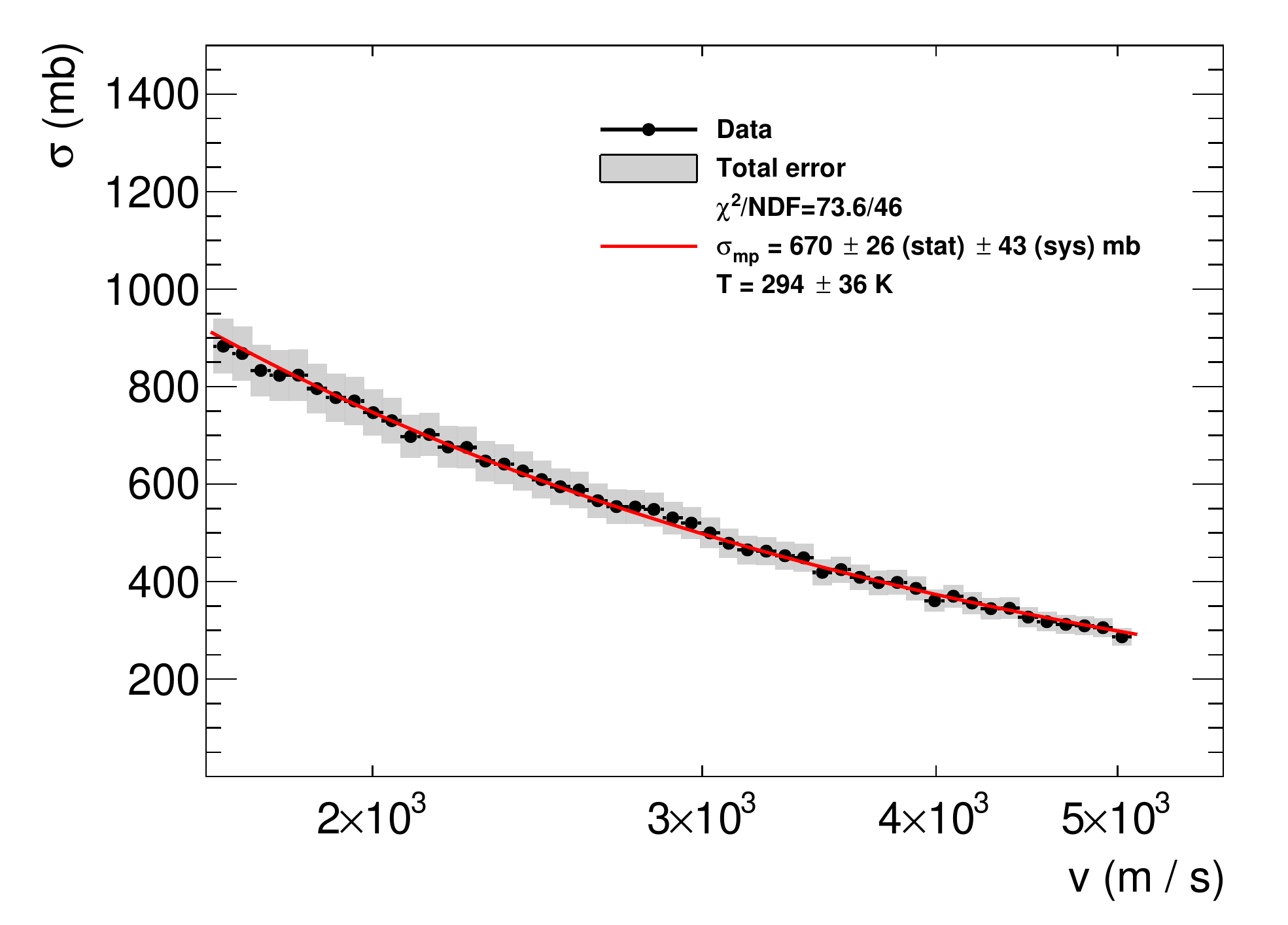}
\caption{$^{40}$Ar cross section as a function of the neutron velocity. 
Both statistical and total errors are shown. 
The fit (red line) estimates both the neutron absorption cross section at the most probable velocity ($\sigma_{mp}$) and the temperature of the moderator ($T$).}
\label{fig:xsec_40Ar_vs_v}
\end{figure}

\begin{table}[tb]
\caption{Summary of the contributions to the final error of the cross section.}
\begin{center}
\begin{tabular}{ccc}
Error & Stat. (\%) & Sys. (\%) \\ 
\hline
$\delta \rho/\rho$ & 0.0 & 0.3 \\
$\delta L/L$ & 0.0 & 2.2 \\
$\delta \varepsilon/\varepsilon$ & 0.3 & 0.9 \\
$\langle \delta G_i /G_i \rangle$ & 2.0 & 0.0 \\
$\langle \delta N_i /N_i \rangle$ & 1.6 & 5.8 \\
\hline
\end{tabular}
\end{center}
\label{tab:xsec_sys_errors}
\end{table}

\begin{figure}[tb]
\centering
\includegraphics[width=\columnwidth]{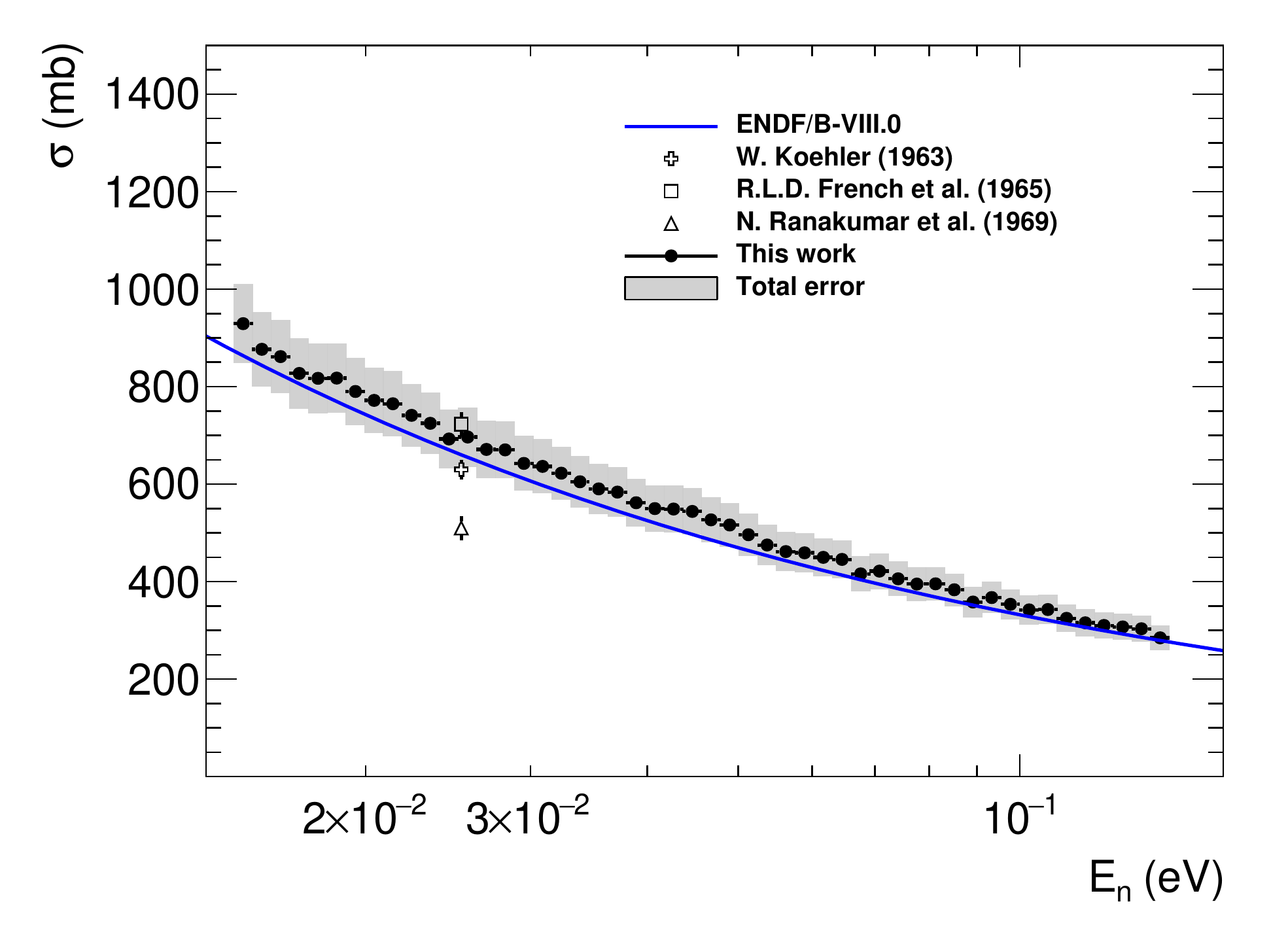}
\caption{Comparison between the results obtained in this work (black circles) and the various measurements and evaluations of the $^{40}$Ar(n,$\gamma$)$^{41}$Ar cross section for thermal neutrons. Each point corresponds to a measurement (box, cross, and triangle) or evaluation (blue line).}
\label{fig:xsec_40Ar_vs_e_after_aced}
\end{figure}

Outside the resonance region ($<15$\,eV for argon, see Ref.~\cite{Brown:2018jhj}), the neutron absorption cross section follows the $1/v$ law~\cite{Westcott:1960effective}:
\begin{equation}
\label{eq:sigmav}
\sigma(v) = \frac{\sigma_{mp} \, v_{mp}}{v}
\end{equation}
where $\sigma_{mp}$ is the neutron absorption cross section at the most probable velocity:
\begin{equation}
v_{mp} = \sqrt{2 k T / m}
\end{equation}
where $k$ is the Boltzmann constant, $T$ is the temperature of the moderator, and $m$ is the reduced mass of the argon-neutron system.

Inside the \textit{i}-th neutron energy bin, the cross section can be evaluated as:
\begin{equation}
\label{eq:sigmai}
\sigma_i = \frac{A}{a_{40} \, \rho \, L \, N_A} \, \frac{G_i}{\varepsilon \, N_{i}} - \zeta_i
\end{equation}
where $A$, $a_{40}$ and $\varepsilon$ are the atomic mass of natural argon, the $^{40}$Ar abundance, and the efficiency to see the $^{41}$Ar gamma cascade after applying the selection cuts, respectively. 
$L$ is the length of the target, and $N_A$ is Avogadro's number.
$G_{i}$ is the number of neutron captures, detected by DANCE, for a given bin. 
$N_{i}$ is the number of neutrons seen by the beam monitor.
Finally, $\zeta_i$ is a theoretical correction to account for the presence of $^{36}$Ar in natural argon.
Natural argon consists of $^{36}$Ar ($0.3336$\,\%), $^{38}$Ar ($0.0629$\,\%), and $^{40}$Ar ($99.6035$\,\%)~\cite{Wang:2014gja}.
We estimate the contribution from $^{38}$Ar to the cross section to be negligible ($<0.1$\,\%), but $^{36}$Ar does require a small correction. 
For this we use both the cross section for $^{36}$Ar$(n,\gamma)^{37}$Ar and the $^{37}$Ar gamma cascade from ENDF/B-VIII.0~\cite{Cameron:2012ogv,Brown:2018jhj}.
Using  $\zeta_i = \frac{a_{36} \, \varepsilon_{36}}{a_{40} \, \varepsilon} ~ \sigma^{36}_i$ gives on average a $1$\,\% correction in Eq.~\ref{eq:sigmai}. 

Fig.~\ref{fig:xsec_40Ar_vs_v} shows the argon cross section as a function of the neutron velocity.
The fit, from $0.015-0.15$\,eV, using Eq.~\ref{eq:sigmav}, yields $T = 294 \pm 36$\,K and $\sigma_{mp} = 670 \pm 26 \text{ (stat.)} \pm 43 \text{ (sys.)}$\,mb.
Tab.~\ref{tab:xsec_sys_errors} summarizes the contributions to the statistical and systematic uncertainties.
$\delta N_i /N_i$ accounts for: the detector stability due to the difference in beam intensity between the argon and calibration runs ($3.1$\,\%), the consistency among the two beam monitors ($2.1$\,\%), and the uncertainties in the sodium calibration procedure ($4.4$\,\%).
The systematic effect of the emission of internal conversion electrons was assessed using BrIcc~\cite{Kibedi:2008etccub} and was found to be negligible ($<0.5$\,\%).

After correcting for the average temperature of the moderator and the target, $\langle T \rangle = 296 \pm 36$\,K, we obtain a result of $\sigma^{2200} = 673 \pm 26 \text{ (stat.)} \pm 59 \text{ (sys.)}$\,mb for the standard thermal cross section.
Moreover, Fig.~\ref{fig:xsec_40Ar_vs_e_after_aced} shows the differential cross section, from $0.015-0.15$\,eV, re-scaled to $300$\,K, alongside historical measurements~\cite{Koehler:196318a,French:1965225,RANAKUMAR1969333} and the evaluation from Ref.~\cite{Brown:2018jhj}.

In conclusion, we have measured for the first time the neutron capture cross section on $^{40}$Ar as a function of energy. 
These results can be used in the design of a new generation of dark matter and neutrino detectors using liquid argon as a detection medium and/or shield.


\section{\label{sec:acknowledgments}ACKNOWLEDGEMENTS}

This work was supported by the U.S. Department of Energy (DOE) Office of Science under award number DE-SC0009999, and by the DOE National Nuclear Security Administration through the Nuclear Science and Security Consortium under award number DE-NA0003180. 
We gratefully acknowledge the logistical and technical support and the access to laboratory infrastructure provided to us by LANSCE and its personnel at the Los Alamos National Laboratory. 
We would also like to thank L. J. Kaufman for providing the target cell. 

\section{Bibliography}

\bibliographystyle{unsrt}
\bibliography{bibliography.bib}

\end{document}